\documentclass[11pt,twoside]{article}
\usepackage{./asp2010}

\resetcounters

\begin{document}

\title{Simulation of microquasars - the challenge of scales}
\author{Rolf Walder,$^1$ Micka\"el Melzani,$^1$  Doris Folini,$^1$ Christophe Winisdoerffer,$^1$ and Jean M. Favre$^2$}
\affil{$^1$\'{E}cole Normale Sup\'{e}rieure, Lyon, CRAL, UMR CNRS 5574, Universit\'{e} de Lyon, France}
\affil{$^2$CSCS Lugano, Switzerland}

\begin{abstract}
  We present first results of a long-term project which aims at
  multi-scale, multi-physics simulations of wind accretion in
  microquasars and high-mass X-ray binaries. The 3D hydrodynamical
  simulations cover all scales, from the circum-binary environment
  down to the immediate vicinity of the black hole. We first introduce
  the numerical method and parallelization strategy of the AMR A-MAZE
  code. We then discuss some preliminary results of how, and on what
  scales, an accretion disk is formed around the black hole. We
  finally present some characteristics of this disk, which is far from
  Keplerian. We emphasize that on all scales shocks play a decisive
  role for the accretion process and the process of structure
  formation -- for the formation of the large scale, nearly coherent
  structure of the disk, but also for the formation of turbulent
  fluctuations.
\end{abstract}

\section{Introduction}

We want to study wind-accreting microquasars (MQ): accreting black
holes (BH) bound to a companion star. This process leads to strong
X-ray emission close to the Eddington limit \citep[for a review,
  see][]{2006ARA&A..44...49R}. This puts MQs into the class of X-ray
binaries. Accretion into a BH is connected to the launch of
relativistic jets, either permanently or episodically (see
\citet{2004ARA&A..42..317F} for a review). This relates MQs to objects
like Gamma-Ray Bursts and Active Galactic Nuclei. Non-thermal
processes, particle acceleration, and associated $\gamma$-ray emission
are observed in MQs \citep[a good review
  is][]{2013A&ARv..21...64D}. The excellent observational coverage and
human accessible time-scales, from milli-seconds to years, make MQs an
ideal laboratory to study multi-scale and multi-physics processes.

Accretion in MQs takes place either by Roche-lobe overflow (RLOF), or
by collecting mass out of the wind of the companion, or by a
combination of both. Main sequence low mass companions have only very
weak winds, which cannot contribute to the accretion into the
BH. Thus, they are pure RLOF systems. Evolved low mass or high mass
companions have strong winds, which can significantly contribute to the
mass feeding of the BH. In many cases, like for instance in the low
mass system GRS 1915+105, hosting a red giant companion, RLOF and
wind-accretion operate in common. Finally, there is a class of systems
where the high mass companion does not fill its Roche lobe and the BH
is fed purely from the companion wind. It is this case which we
discuss in this paper.

For our simulations and the results presented here, we take the
parameters of the well observed system Cyg~X-1
\citep{2011ApJ...742...84O}. A major issue in wind-accreting systems
is how winds are accelerated in the presence of a second ionizing
source, the X-rays originating from the vicinity of the BH
\citep{1990ApJ...365..321S, 1994ApJ...435..756B,
  2012A&A...542A..42H}. To keep this first multi-scale study of MQs
simple, we decided to ignore this question. Instead, we performed a
set of simulations with different constant values for the wind speed:
$750$~km/s, $850$~km/s, $1000$~km/s, $1500$~km/s, $2000$~km/s, and
$2500$~km/s. We launch the wind with this velocity directly from the
surface of the companion star. Another major issue is that radiative
transfer effects influence the thermodynamical state of the flow, and
subsequently its dynamics. For this first study, we us a simple
polytropic equation of state with different indices: $\gamma =5/3,
4/3, 1.1, 1.01$. An asset of our simulations is, by contrast, their
spatial resolution, from the circum-binary scale (computational domain
: $10^{14}$~cm cubed) down to some 10 gravitational radii of the BH
($R_G = 2 G M_{BH}/ c^2 = 4.4 \cdot 10^{6}$~cm). This allows, for the
first time, to study the accretion wake and the formation of an
accretion disk under otherwise idealized conditions. Euler equations
are solved. More complex physics and the acceleration process of the
wind will be included in future. Also, in this paper, we strictly
concentrate on the case of the $750$~km/s wind and $\gamma = 1.1$.
%, corresponding to strong energy loss by radiation.

In Sect.~\ref{Sec:NumericalMethods}, we present our numerical methods,
in Sect.~\ref{Sec:Results} our results, and we conclude in
Sect.~\ref{Sec:Conclusions}.

\section{Numerical method}
\label{Sec:NumericalMethods}

The simulations were performed with the A-MAZE simulation toolkit
\citep{2000ASPC..204..281W, 2003ASPC..288..433F, 2013A&A...558A.133M},
comprising 3D parallelized MHD, radiative transfer, and particle in
cell plasma codes. For the hydrodynamical simulations we used the
explicit multi-dimensional method by~\citet{colella:90}, the Riemann
solver by~\citet{Colella-Glaz:85}, and the block-structured AMR
algorithm by~\citet{berger:85}.

\begin{figure}[tbp]
\centerline{
  \includegraphics[width=5.3cm]{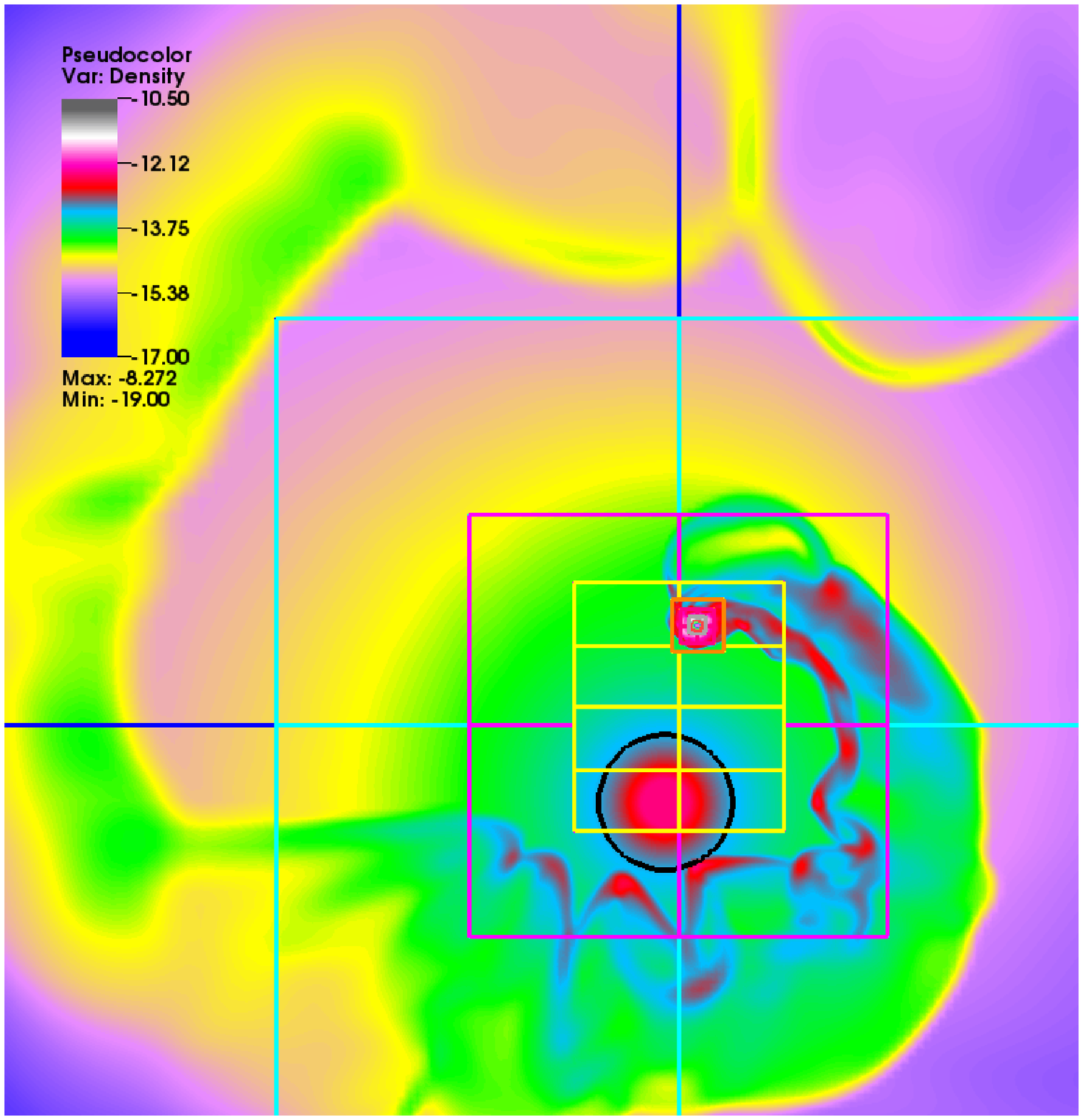}
  \includegraphics[width=5.3cm]{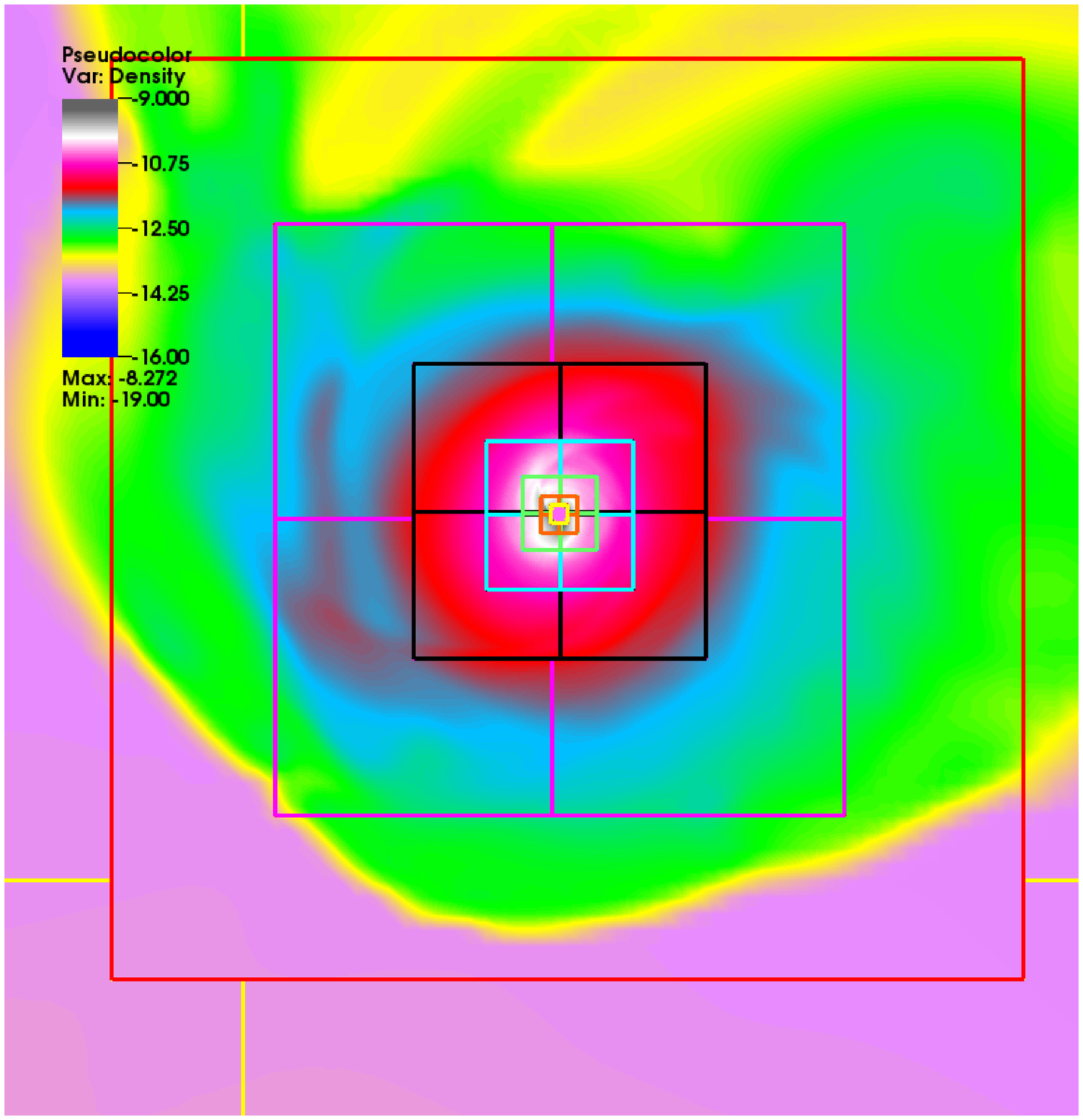}
         }
         \caption{Structure on the circum-binary scale (left panel,
           $1.5 \cdot 10^{13}$~cm, approximately 1/6 of the
           computational domain) and on the scale of the bow-shock
           heading the accretion wake (right panel, $2-3 \cdot
           10^{11}$~cm, or approximately $45'000-70'000$~R$_G$ from
           the BH, the two numbers indicate the fluctuation
           size). 2D-extracts of density in the orbital plane are
           shown. The mesh structure with its different levels of
           refinement is superimposed. The finest level in this case 
           is 14, each level refining by a factor of 2.  The radius of
           the accreting sphere (BH boundary) corresponds to $150$~R$_G$.}
\label{Fig:Large-Disk-ScaleDensityMesh}
\end{figure}

Our adaptive grid is setup similarly as in \citet{2008A&A...484L...9W}
(see also Fig.~\ref{Fig:Large-Disk-ScaleDensityMesh}). The orbital
scale is covered by finer meshes than the circum-binary scale. A
nested tree of grids is constructed around the accreting BH moving
through the wind of the companion. It is only at this single spot,
where many levels of refinement (up to 19 for certain models) are
needed to resolve the process of disk formation and to capture the
scale of the gravitational radius of the BH. Note that in spatially
refined blocks the time-step is adapted as well, such that for all
grids the cfl-number can be kept constant. In this way, we are able to
resolve time-scales associated to each region in an optimal way. Time
steps in the vicinity of the BH are milli-seconds, corresponding to
dynamical time-scales in this region (e.g. quasi-periodic
oscillations, QPOs). Similarly, time-steps on the orbital scales are of
order of a minute. For one orbit of 5.6 days, we need in this way
around 10'000 time-steps on the coarsest grid.

Nevertheless, although the code shows very good weak scaling even for
fully dynamic AMR if blocks of more than $30^3$ cells are used
(scaling limit), the setup used here limits the number of processors
one can use. In many simulations, fine grids are needed throughout the
computational domain, thus dominate the computational costs. Using
many processors then reduces the wallclock time if parallelization on
the finest level scales optimal - even if coarser levels have some
scaling flaws due to too small blocks.  However, for the accretion
problem, the size (measured in cells) of refined regions is about
constant for all levels.  The scaling limit thus provides a natural
limit of parallelization that could only be overcome by improving
strong scalability.  For this study, we used 8 processors working on 8
blocks of $40^3$ cells, resulting in a mesh of about 7 million (14
refinement levels), or 10 million cells (19 refinement levels),
respectively. Nevertheless, thanks to larger time-steps on coarser
meshes, the net gain in both, CPU and wallclock time is
considerable. Scalar computation would result in 8 times more
wallclock, non-time adaptivity in much more CPU time to achieve the
same result. In a later stage of the study, we may increase the size
of refined regions and use 64 processors. Data-files in hdf5 format
take 1 to 1.5 GB for one time step.

A final note on numerical methods: the presented problem is hard to
resolve on the basis of an explicit solver as time-scales involved
span milli-seconds to years. Part of the problem can be overcome by
time-step adaption as described above, but the problem remains rather
stiff.  Implicit solvers could possibly overcome the stiffness,
resulting in a net gain of compute time and diminishing use of compute
resources. We have started to develop such solvers
\citep{2011A&A...531A..86V, 2013A&A...555A..81V}, which indeed prove
to be much faster than explicit solvers. However, these solvers are
not yet ready to treat also shock waves, and the solvers are not yet
implemented into the AMR code. But these issues should not be
principle obstacles for the use of implicit solvers, see
\citet{birken_habilthesis} for a review of the newest developments in
the field. On the other hand, implicit-explicit schemes may turn out
to be an interesting alternative to fully implicit schemes
\citep[e.g.][see also the contribution of F.~Kupka in this
  volume]{2012JCoPh.231.3561K}.

\begin{figure}[tbp]
\centerline{
 \includegraphics[width=12.cm]{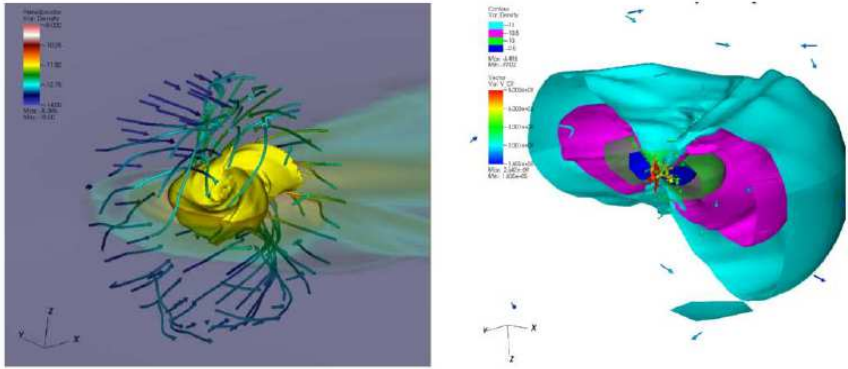}
         }
         \caption{Behind the bow shock, on a scale of $R \approx
           50'000$~R$_G$, a spinning structure starts to form (left
           panel). Shown is density in orbital plane, a 3D density
           isosurface and the velocity field in streamlines. It is
           only on a much smaller scale, on a scale of about
           $5'000$~R$_G$, where the spinning structure is flattened
           to form a disk (right panel). Shown here are different
           density isosurfaces and the flow field in vectors.}
\label{Fig:Disk_Formation}
\end{figure}
\section{The $v_{W} = 750$~km/s multi-scale model: first results }
\label{Sec:Results}

We briefly describe the process driving the nearly spherically
symmetric and homogeneous wind of the companion to form an accretion
cone, then a subsequent spinning structure around the BH, and finally
a non-classical accretion disk close to the BH, at the same time
developing turbulent fluctuations on each of these scales.

The large scale, circum-binary structure of the system is dominated by
the spirally-shaped accretion-cone trailing the bow-shock around the
black hole (Fig.~\ref{Fig:Large-Disk-ScaleDensityMesh}, left
panel). Though only the part very close to the BH is involved in the
accretion process, the enhanced density of this cone may lead to
non-negligible absorption processes observable in the binary spectrum,
causing either secondary dimming or even occultation of the X-ray
source \citep{2000A&A...354.1014D, 2004AJ....127.2310G}.

Closer to the BH, to about half the Bondi-Hoyle accretion radius or
$45'000-70'000$~R$_G$, a bow-shock forms in front of the BH moving on
its orbit through the wind
(Fig.~\ref{Fig:Large-Disk-ScaleDensityMesh}, right panel). The bow
shock is highly variable in time and space, mostly due to turbulent
fluctuations in the wake. The shape-variations of the shock itself
force the turbulence \citep{foglizzo:02, 2006A&A...459....1F}. The
shocked material is immediately re-accelerated by the gravitational
field of the BH. It collides with other material falling in the
accretion cone towards the BH. The net angular momentum of these
converging flows lets arise a structure that spins pro-grade around
the BH. We emphasize that a substantial part of the angular momentum
is dissipated in the many shocks that accompany the merging of these
different colliding flows. The spinning structure on this scale bares
no resembling with a disk. It is still an essentially 3D structure
interwoven with strongly varying shocks and showing strong
discontinuities in all flow variables (Fig.~\ref{Fig:Disk_Formation},
left). Shocks play an important role in the transport of angular
momentum down to the BH scale.

In each shock passage, the falling material looses angular
momentum. This causes a flattening and circularizing of the
flow. Shocks become steadily more coherent. In the inner part, shocks
are 1- or 2-armed spirally-shaped waves, whose positions only evolve on
secular time scales. In this hydrodynamical model, the emerging disk is
neither thin nor Keplerian (Fig.~\ref{Fig:Disk_Formation}, right), but
nevertheless a disk which may emit not too differently from a
Keplerian disk. The heating process, however, is dominated by shocks
rather than viscous shear, though this second process certainly also
plays a role. Density and velocity fluctuations within the disk are
still of order 20--30 percent. The thickness of the disk is
essentially given by the pressure of turbulent fluctuations within the
disk and the kinetic pressure built up by the material steadily
raining down on the disk from flow features above and below the
disk. In some cases, shocks in the down-draining material can build
up, confining the disk. Both, turbulent fluctuations and the
down-drain of material, may cause flickering of the emission as
observed in many wind-accreting X-ray binaries.

\begin{figure}[tbp]
\centerline{
  \includegraphics[width=12.8cm]{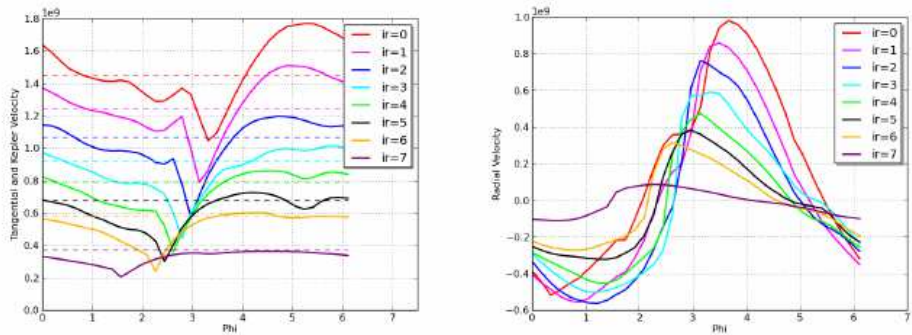}
         }
         \caption{Tangential (left panel, together with Kepler
           velocity) and radial component (right panel) of the
           velocity [cm/s] in the central plane of the disk, shown for
           different radii of the disk (214~$R_G$ (ir=0), 290~$R_G$,
           393~$R_G$, 531~$R_G$, 720~$R_G$, 976~$R_G$, 1322~$R_G$,
           3286~$R_G$ (ir=7)). Positive tangential velocity indicates
           a pro-grade rotation of the disk (i.e. particles move from
           small to large $\phi$). Positive radial velocity indicates
           a movement towards the BH. The spiral shock wave is visible
           as minima of the tangential velocity / maxima of the radial
           velocity, shifting with the radius. Velocity components
           normal to the disk have values of the same order.}
\label{Fig:Tangential_KeplerOrbit}
\end{figure}

Fig.~\ref{Fig:Disk_Formation}, right, demonstrates that the disk is
not symmetric, but has a larger scale-height in its right part. Also,
the disk is occasionally warped and/or tilted against the orbital
plane. However, we cannot yet exclude that this is an artifact of
non-relaxed initial conditions. Fig.~\ref{Fig:Tangential_KeplerOrbit}
illustrates both, the velocity variations within the disk (along
imaginary circular orbits for different radii) and angular momentum
advection by spiral shocks. Clearly, flow motion is not at all along
circular orbits, as indicated by the strong smooth variation of the
tangential and radial velocity components. The velocity components
also change abruptly. Each shock passage decelerates the fluid parcel,
breaking the tangential flow component, the centrifugal force is
diminished, gravity acts, acceleration towards the BH occurs. In
Fig.~\ref{Fig:Tangential_KeplerOrbit}, the spiral shock is visible as
points of smallest tangential velocities and largest radial
velocities, shifting with radius.

\section{Discussion and Conclusions}
\label{Sec:Conclusions}

We presented first results of our project on multi-scale,
multi-physics simulations of microquasars. We first discussed the
efficiency of parallel computation of time-adaptive AMR simulations
using many levels of refinement at one spatial spot only. We then
presented a preliminary analysis of how a disk is formed out of the 3D
flow accreting material from a wind shed by the companion star. We
note that such a disk is not always formed. Depending on the wind
speed of the companion, the 3D structure of the accretion flow may be
essentially conserved down to the scale of the gravitational radius of
the black hole. Angular momentum is advected by a network of shocks.
This is similar to what has been discussed
by~\citet{2010ASPC..429..173W} for the low mass binary case, where we
called this case the accretion ball regime \citep[see
  also][]{2005ARep...49..884M}. For very high speed winds, we find
stable disks which are rotating retrograde. A more thorough discussion
of these other accretion regimes will be published elsewhere.

As perspective we note that a similar study is in preparation for the
MHD case. With this forth-coming study, we hope to learn more on
whether magnetic fields essentially change the disk-formation
process. Certainly one expects another essential heating process to
be present, magnetic reconnection. On a longer time scale, we want to
combine MHD simulations with kinetic simulations and radiative
transfer.

\acknowledgements{We acknowledge support from the French Stellar
  Astrophysics Program PNPS and computing time from the Grand
  Equipement National de Calcul Intensif (GENCI), project number
  x2012046960 and support from the P\^ole Scientifique de Mod\'elisation
  Num\'erique (PSMN) at ENS-Lyon.}

\bibliography{Walder_X-Ray-Binaries}

\end{document}